\documentclass{article}
\usepackage{spconf,amsmath,graphicx}
\usepackage{kotex, xcolor}
\usepackage{amssymb}

\usepackage{booktabs}       

\usepackage{multicol}
\usepackage{multirow}

\usepackage[absolute,showboxes]{textpos}

\TPMargin{5pt}

\newcommand{\copyrightline}{
    \begin{textblock}{0.84}(0.08,0.93)    
         \noindent
         \footnotesize
         \copyright 2021 IEEE. Personal use of this material is permitted. Permission from IEEE must be obtained for all other uses, in any current or future media, including reprinting/republishing this material for advertising or promotional purposes, creating new collective works, for resale or redistribution to servers or lists, or reuse of any copyrighted component of this work in other works.
    \end{textblock}
}

\newcommand{\R}{\mathbb{R}}

\title{SubSpectral Normalization for Neural Audio Data Processing}
\name{\begin{tabular}{c}Simyung Chang \quad Hyoungwoo Park \quad Janghoon Cho \quad Hyunsin Park \\ \textit{Sungrack Yun \quad Kyuwoong Hwang} \thanks{  ${}^{\dagger}$ Qualcomm AI Research is an initiative of Qualcomm Technologies, Inc.} \end{tabular}}

\address{Qualcomm AI Research${}^{\dagger}$, Qualcomm Korea YH\\
{\small\texttt {\{simychan, hwoopark, janghoon, hyunsinp, sungrack, kyuwoong\}@qti.qualcomm.com} }}

\begin{document}
%
\maketitle
\copyrightline
\begin{abstract}
Convolutional Neural Networks are widely used in various machine learning domains. In image processing, the features can be obtained by applying 2D convolution to all spatial dimensions of the input. However, in the audio case, frequency domain input like Mel-Spectrogram has different and unique characteristics in the frequency dimension. Thus, there is a need for a method that allows the 2D convolution layer to handle the frequency dimension differently. In this work, we introduce \textit{SubSpectral Normalization} (SSN), which splits the input frequency dimension into several groups (sub-bands) and performs a different normalization for each group. SSN also includes an affine transformation that can be applied to each group. Our method removes the inter-frequency deflection while the network learns a frequency-aware characteristic. In the experiments with audio data, we observed that SSN can efficiently improve the network's performance. 
\end{abstract}
\section{Introduction}
\label{sec:intro}
The Convolutional Neural Networks (CNNs) have been widely used in the recent studies on deep neural networks for various domains such as image, audio, and text. Early researches on CNNs have been mainly studied in the image domain, and enormous improvements and achievements are obtained with some architectures, VGG~\cite{SimonyanZ14a} or ResNet~\cite{he2016deep}, for many computer vision tasks. These research results have been applied to audio and speech tasks with various modifications in the architectures \cite{salamon2017deep, tang2018deep, choi2019temporal}. Most methods \cite{abdel2014convolutional, hori2017advances, 7952132, koutini2019receptive} based on the frequency domain feature (e.g. Mel-Spectrogram) use the architecture consisting of multiple 2D convolution layers.   

The 2D convolution operation equally processes the input data in vertical and horizontal directions. As illustrated in Figure 1(a), this processing is proper for image-domain tasks to extract the features of objects placed in different locations given an image. However, the audio feature, Mel-spectrogram in Figure 1(b), shows some unique characteristics depending on the frequency dimension (vertical direction). Thus, the same 2D convolution operation used in image-domain tasks may not be
appropriate for the audio-domain tasks. Several studies have been reported to address this problem and propose an architecture where separate convolution layers are designed for each frequency sub-band \cite{phaye2019subspectralnet, mcdonnell2020acoustic, kao2019sub}. However, this leads to much computation and memory with the increase of the number of sub-bands. 
It's hard to apply this architecture to other applications since it's designed for a specific task.

\begin{figure}[t]
\centering
\includegraphics[width=0.9\linewidth]{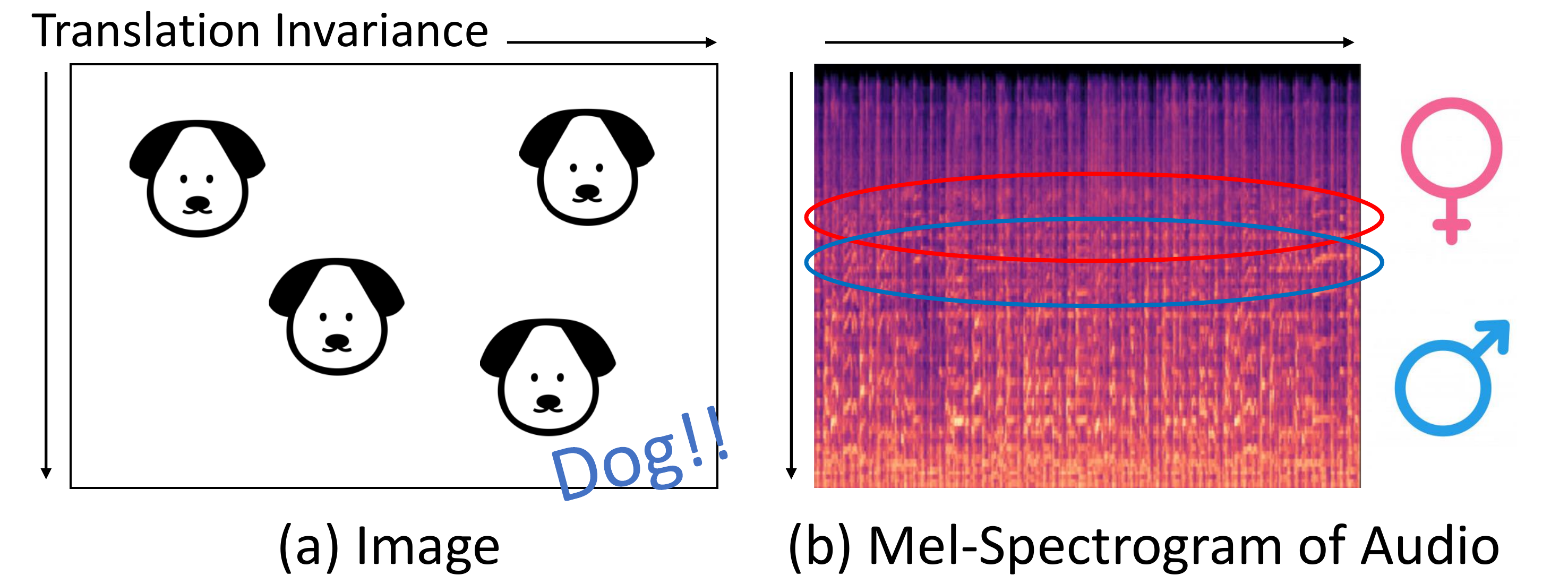}
\vskip -0.07in
\caption{\textbf{2D convolution on image and audio input} Unlike image processing, the feature in different audio frequency bands has different information.}
\label{fig:problem}
\end{figure}

To handle these problems, we consider a normalization layer commonly used in CNN. Batch normalization, one of the most widely used normalization methods, uses batch statistics to normalize each channel. 
But the normalization is equally performed in the frequency and temporal direction. Thus, it may not be easy to interpret the unique characteristics of each frequency band differently. Furthermore, if there is an imbalance of the scale in data, this is also kept in the normalized feature. To overcome these limitations, we propose a novel normalization technique, SubSpectral Normalization (SSN). Our method divides the frequency dimension into several sub-bands and normalizes each sub-band. By applying SSN, each band's scale imbalance can be adjusted. The convolution kernel for each band acts as a different filter by performing other affine transformations for each group.

\begin{figure*}[t]
\centering
\includegraphics[width=0.85\linewidth]{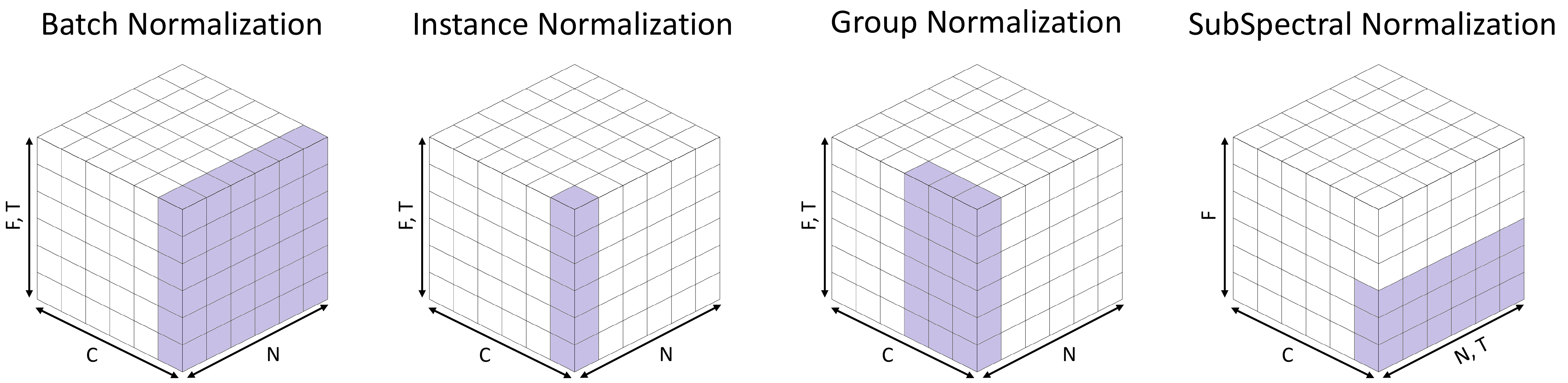}
\vskip -0.05in
\caption{Normalization methods on Frequency-Time audio input, with N as batch axis, C as channels, F as frequency and T as time axis. SSN shows the case of two sub-bands.}
\label{fig:ssn}
\end{figure*}
We applied our method on two different tasks to confirm SSN's effectiveness: acoustic scene classification and keyword spotting. SSN can replace batch normalization layers of the models without increasing computation. The experimental results show that our method could significantly improve model performance by changing the existing normalization layer.

Our contributions are summarized as follows:\\
(1) We propose SubSpectral Normalization (SSN), which splits the frequency dimension into multiple sub-bands and normalizes each group. \\
(2) SSN can normalize each sub-frequency band and allows a convolution filter to behave like multiple filters with only small additional parameters. \\
(3) SSN can improve performance by just replacing the normalization layer of the model.

\section{Related Works}
\label{sec:related}
\textbf{Normalization.} 
Many normalization methods have been proposed in deep neural networks. Batch normalization (BN) \cite{ioffe2015batch} operates normalization along the batch dimension. 
Some recent studies do not compute along batch dimensions to overcome the drawbacks of using batch statistics.
Layer normalization \cite{ba2016layer} operates along the channel dimension to improve performance in small mini-batch size in the recurrent neural networks (RNNs). Instance normalization (IN) \cite{ulyanov2016instance} normalizes each channel independently and applies it at test time and training. Group normalization (GN) \cite{wu2018group} proposes group-wise computation along the channel axis to solve the degradation of performance because of dependency on the batch size in case of small mini-batch size. Weight normalization (WN) \cite{salimans2016weight} performs normalization for the filter weights.
Despite these various studies on normalization, the previous methods still equally normalize all features of the same channel.
Different from previous normalization methods, we propose subspectral normalization (SSN) that performs along the sub-bands of frequency dimension. It is similar to apply a different convolution filter at each sub-band in spectrogram.

\noindent\textbf{Using sub-frequency bands.}
SubSpectalNet \cite{phaye2019subspectralnet} trains separate CNNs on sub-spectrograms divided along the frequency axis from the spectrogram, and each CNN learns properties from different frequency bands. Mcdonell \textit{et al.} \cite{mcdonnell2020acoustic} apply two parallel paths for high and low frequencies and combine the two paths using late fusion along frequency axes. Sub-band CNN  \cite{kao2019sub} splits spectrogram into overlapped sub-bands and concatenates the different features that are extracted from each sub-band after the first convolutional layer.
In this paper, we reconsider the normalization layer to handle the frequency band differently. Our method requires less additional computation and has little effect on the model size. SSN can be applied to conventional CNN models by replacing the BN layer.  

\begin{figure}[t]
\centering
\includegraphics[width=1.\linewidth]{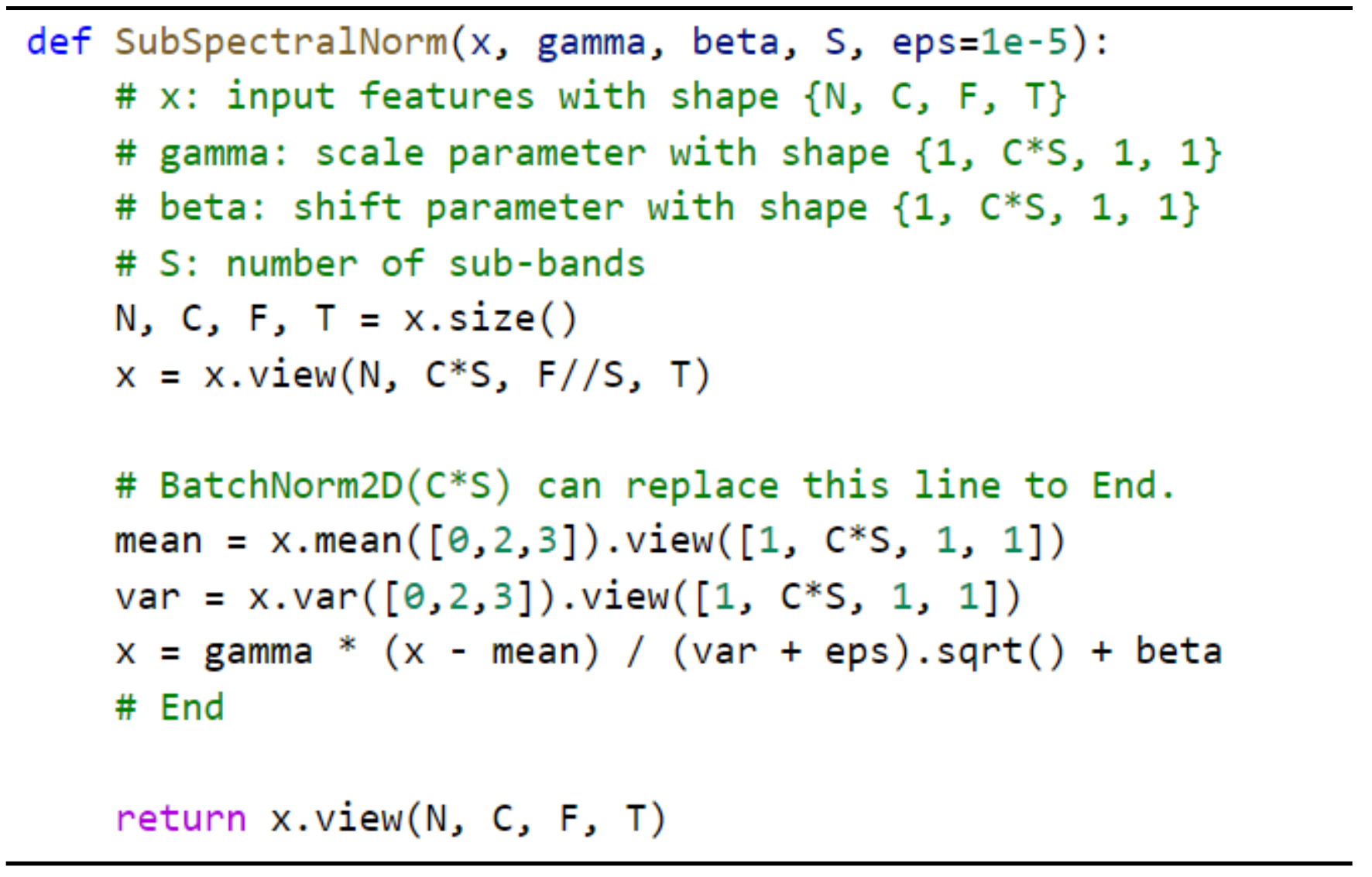}
\vskip -0.07in
\caption{PyTorch code of SubSpectral Normalization with affine transormation type \textit{Sub}} 
\label{fig:code}
\end{figure}

\section{SubSpectral Normalization}
\label{sec:method}
In this section, we present a novel normalization method, SubSpectral Normalization (SSN), which can be applied to audio-domain tasks based on 2D convolutional networks. Our method splits the input frequency dimension into several groups (sub-bands) and performs a different normalization for each group. Figure \ref{fig:ssn} shows the comparison of conventional normalization methods with SSN.

Normalization methods can be expressed as follows:
\begin{equation}
    \tilde{x} = \frac{1}{\sigma} (x - \mu).
\label{eq:norm}
\end{equation}

\noindent Here, $x$ denotes the input feature, and $\mu$ and $\sigma$ are the mean and standard deviation of the $x$, respectively.
In Batch Normalization (BN), $x$ is a feature of the same channel in a mini-batch, and $\mu$ and $\sigma$ denote the mean and standard deviation of this feature x. 
For the SSN, we divide the frequency dimension into multiple groups, and $x$ represents one sub-band of these groups, not the entire feature of one channel. $\mu$ and $\sigma$ are also calculated for each sub-band.
Figure \ref{fig:code} is the code that implements a training mode of SSN on PyTorch.  As shown in the code, SSN can be performed by separately applying Batch Normalization to each sub-band. And the frequency groups are divided equally for efficient computation. SSN gives the effect that the parameters of the following convolution layer are defined differently for each sub-band.

When the number of sub-bands is $S$ and $i$ denotes the $i$th sub-band, 
the normalized feature $\tilde{x_i}$ of the sub-band feature $x_i$ can be defined as:
\begin{equation}
    \tilde{x_i} = \gamma^{SSN}\cdot\frac{1}{\sigma_i} (x_i - \mu_i) + \beta^{SSN},
\label{eq:ssnorm_all}
\end{equation}
where $\mu_i$ and $\sigma_i$ are the mean and standard deviation for the $i$th sub-band. $\gamma^{SSN}$ and $\beta^{SSN}$ denote scale and shift parameters of SSN, respectively. Here, SSN's affine transformation parameters are shared by the entire frequency dimension, not each sub-band. We define this transform type as \textit{All}. The SSN can perform separate affine transformation for each sub-band, which is defined as follows:

\begin{equation}
    \tilde{x_i} = \gamma^{SSN}_i\cdot\frac{1}{\sigma_i} (x_i - \mu_i) + \beta^{SSN}_i,
\label{eq:ssnorm_sub}
\end{equation}
where $\gamma^{SSN}_i$ and $\beta^{SSN}_i$ are scale and shift parameters for the $i$th sub-bands. We define this transformation type as \textit{Sub}.

If there is a convolution layer following SSN, the merged parameter of two layers for each sub-band can be defined as follows:
\begin{equation}
    W^{conv}_i = \gamma^{SSN}_i \cdot W^{conv},
\label{eq:conv_w}
\end{equation}
and
\begin{equation}
    B^{conv}_i = \gamma^{SSN}_i \cdot B^{conv} + \beta^{SSN}_i.
\label{eq:conv_b}
\end{equation}
Here, $W^{conv}\in \R^{C \times (C^{prev} \cdot k^2)}$ and $B^{conv}\in \R^{C}$ denote the weight and bias of the next convolution layer with $k\times k$ size kernels, where $C^{prev}$ and $C$ are the number of input channels and output channels, respectively.
Using SSN instead of BN, the next convolution layer for the $i$th sub-band is defined as a function of $W^{conv}$, $B^{conv}$, $\gamma^{SSN}_i$ and $\beta^{SSN}_i$. It means that the convolution with SSN can operate differently on each sub-band compared to the convolution with BN, which works equally on the whole frequency dimension.

When applying SSN to CNNs, the user can control the number of sub-bands and the type of affine transformation as hyper-parameters, and we denote it as SSN($S$=\textit{number of sub-bands}, $A$=\textit{affine type}) in this paper. To this, SSN($S$=1, $A$=\textit{All}), SSN($S$=1, $A$=\textit{Sub}) and BN are equivalent operations.


\begin{table}[t]
\caption{Results on TAU Urban Acoustic Scenes 2019.
}
\vskip 0.1in
\label{tab:dcase2019}
\resizebox{1.\linewidth}{!}{
\begin{tabular}{lcc}
\toprule
Model               & Accuracy & \#Params \\
\midrule             
CP-ResNet(ch64) w/ BN & 82.3\% $\pm 0.19$ & 899K \\
CP-ResNet(ch64) w/ BN + Input Norm & 82.7\% $\pm 0.35$ & 899K \\
CP-ResNet(ch128) w/ BN & 83.2\% $\pm 0.22$ & 3,567K \\
\midrule
\textbf{CP-ResNet(ch64) w/ SSN}(S=2, A=Sub)        & \textbf{83.6\%} $\pm 0.07$ & 907K \\
\textbf{CP-ResNet(ch128) w/ SSN}(S=2, A=Sub)       & \textbf{84.1\%} $\pm 0.20$ & 3,583K \\
\bottomrule
\end{tabular}
}
\end{table}

\begin{figure}[t]
\centering
\includegraphics[width=0.8\linewidth]{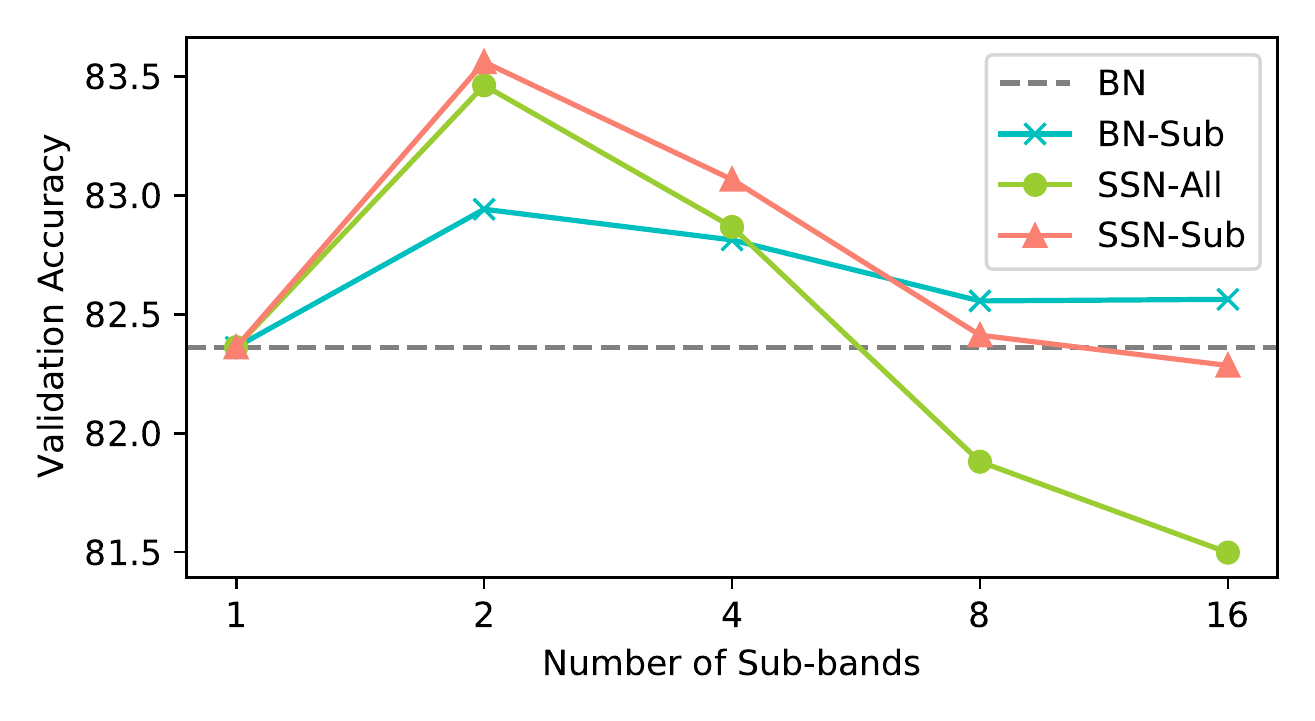}
\vskip -0.1in
\caption{Validation accuracy of CP-ResNet (ch64) depending on the hyper-parameters of SubSpectral Normalization. Each accuracy denotes an average of 5 runs. \textit{BN-Sub} denotes BN with sub-bands affine transformation.}
\label{fig:ablation}
\end{figure}

\section{Experiments}
\label{sec:exp}
We have experimented with our method on two different tasks. One is an acoustic scene classification, and the other is keyword spotting. In the following experiments, we demonstrate the potential of SubSpectral Normalization (SSN) by applying it to audio tasks dealing with ambient sound and speech data.

\subsection{Experimental Setup}
\textbf{Acoustic Scene Classification}
We evaluate SSN using the TAU Urban Acoustic Scenes 2019 10 class dataset \cite{Mesaros2018_DCASE} which consists of acoustic scene samples recorded in 12 different European cities. Each recording has the audio scene label (one of 10 scenes: e.g., `airport' or `shopping mall'). For the task 1A, the dataset contains the ten acoustic scenes, and the development set includes 40 hours of data with 14,400 segments. In the experiments, we select 9,185 segments and 4,185 segments for the training and evaluation dataset, respectively: we use the split in the first fold of the validation set.

\noindent\textbf{Keyword Spotting}
We select the google speech command dataset \cite{sainath2015convolutional} to evaludate SSN on speech data. The dataset has 65,000 one-second long utterances of 30 short words, by thousands of different people. Following Google’s implementation, we distinguish 12 classes: \textit{yes}, \textit{no}, \textit{up}, \textit{down}, \textit{left}, \textit{right}, \textit{on}, \textit{off}, \textit{stop}, \textit{go}, silence and unknown. The utterances were then randomly split into training, development, and evaluation sets in the ratio of 80:10:10, respectively.

\subsection{Acoustic Scene Classification}
In this section, we conduct experiments using TAU Urban Acoustic Scenes 2019 dataset. We select CP-ResNet~\cite{koutini2019receptive} as the baseline, which shows high performance with simple ResNet architecture. It uses 256 bins Mel-Spectrogram as input, and the setting of the experiment follows \cite{koutini2019receptive}.
Table~\ref{tab:dcase2019} shows the performance when we applied SSN to the baseline models. By applying SSN to CP-ResNet (ch64) with 64 base channels, we got an accuracy improvement of 1.3\%. It is higher than CP-ResNet (ch128), which is four times bigger model. \textit{Input Norm} is the result of normalizing the input Spectrogram by all frequency bins. This result shows that just normalizing the input cannot reach the same effect as SSN. We also applied SSN to a bigger model, CP-ResNet (ch128), and obtained a 0.9\% accuracy gain. This consistent improvement shows that SSN works very effectively in acoustic scene classification. We obtained the best results when the number of sub-bands $S$ is two, and affine transformation is applied separately for each sub-band.

Figure \ref{fig:ablation} shows the validation accuracy according to hyper-parameters of SSN. SSN shows better performance when the individual affine transformation is performed (\textit{SSN-Sub}) than when applied to the whole frequency dimension (\textit{SSN-All}). And when the number of sub-bands is between 2 and 4, SSN performs quite better than BN. But performance decreases as the number of sub-bands increases. When applying the sub-bands affine transformation to BN (\textit{BN-Sub}), there is a slight accuracy improvement, but it is quite lower than \textit{SSN-Sub}. 
These results show that proper sub-band size is more important than eliminating all frequency bin's characteristics.

\begin{table}[t]
\caption{Results on Google Speech Command dataset. The numbers marked with $^*$ are taken from each paper, and $\ddag$ denotes the result of training the same epoch as \cite{choi2019temporal}.}
\vskip 0.1in
\label{tab:speech}
\resizebox{1.\linewidth}{!}{
\begin{tabular}{l@{\hskip 0.6in}c@{\hskip 0.3in}c}
\toprule
Model               & Test Accuracy & \#Params \\
\midrule               
res8 w/ BN~\cite{tang2018deep} & 94.1\% $\pm 0.35$  $^*$& 111K \\
res15 w/ BN~\cite{tang2018deep} & 95.8\% $\pm 0.48$  $^*$& 239K \\
TC-ResNet14-1.5~\cite{choi2019temporal} & 96.6\% $^*$  & 305K \\
EdgeSpeechNet-A~\cite{lin2018edgespeechnets}  & 96.8\% $^*$ & 107K \\
\midrule
\textbf{res8 w/ SSN}($S$=4, $A$=\textit{Sub})   & 95.4\% $\pm 0.22$ & 113K \\
\textbf{res15 w/ SSN}($S$=4, $A$=\textit{Sub})  & 96.8\% $\pm 0.13$ & 243K \\
\textbf{res15 w/ SSN}($S$=4, $A$=\textit{Sub}) \ddag & \textbf{97.5\%} $\pm 0.15$ & 243K \\
\bottomrule
\end{tabular}
}
\end{table}

\subsection{Keyword Spotting}

To verify our method on speech data, we evaluate SSN on the baseline \cite{tang2018deep}, which has multiple 2D convolution layers with a residual architecture. The baseline receives MFCC of 40 features with a window size of 30ms and a hope size of 10ms.
We conduct experiments by replacing BN with SSN, and Table~\ref{tab:speech} shows the result. Unlike in acoustic scene classification, SSN shows the best results with $S$ of 4 in these experiments. There was a notable increase in accuracy, with a small parameter increase within 2\%. SSN showed a bigger performance improvement on \textit{res8} than the large model \textit{res15}, which already has high accuracy. Even though \textit{res15} has a very simple structure consisting of several residual blocks, \textit{res15~w/~SSN} shows similar performance to the recent models \cite{choi2019temporal, lin2018edgespeechnets}. By using the same training budget with \cite{choi2019temporal}, \textit{res15~w/~SSN} shows the state of the art performance among the methods that do not use any additional noise or data. These results show that the SSN allows better processing of audio input even in simple structured models.

\subsection{Qualitative Analysis}
\begin{figure}[t]
\centering
\includegraphics[width=1.0\linewidth]{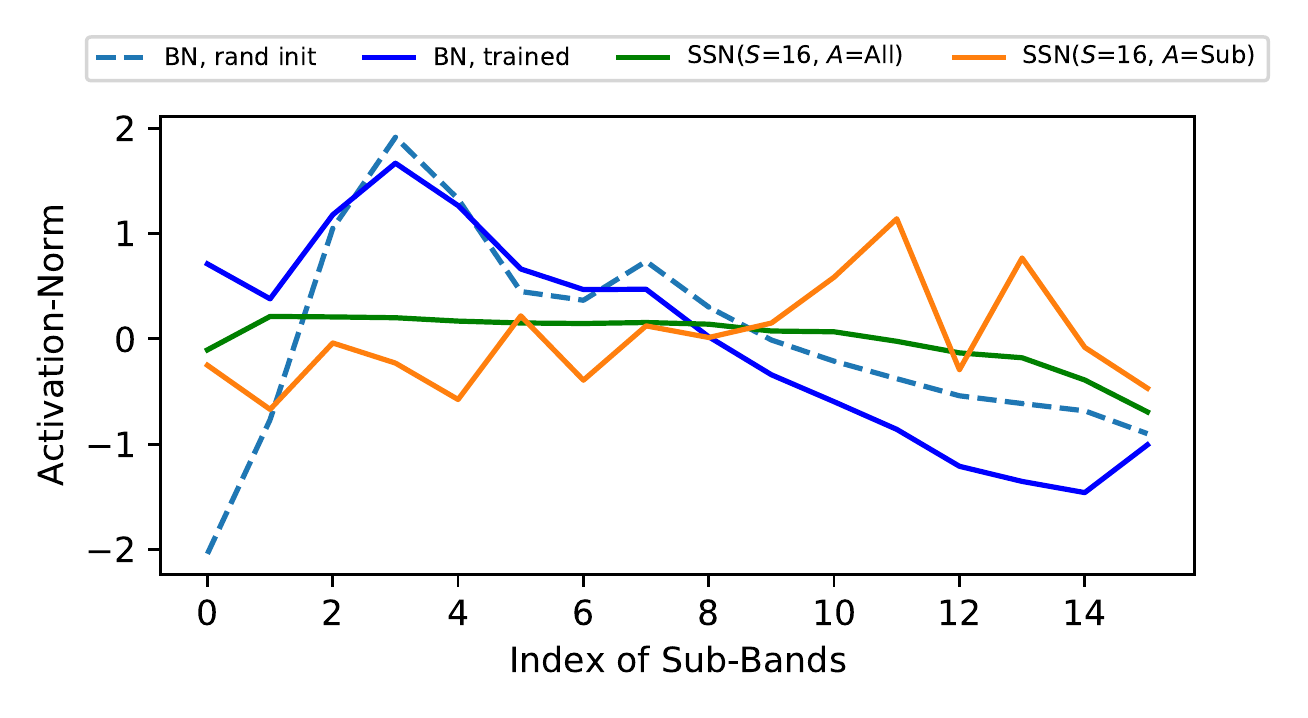}
\vskip -0.2in
\caption{Comparison of activation-norm according to normalization methods in CP-ResNet(ch64). 
}
\label{fig:actnorm}
\end{figure}

We check how each frequency bin changes when applying SSN to confirm the effect of SSN. Figure~\ref{fig:actnorm} shows the scale of the activation through the convolution layer for each frequency bin. We have obtained an activation scale with the L1 norm and averaged it for each sub-band. We normalize each activation scale with zero mean and unit variance to compare each method. After the model is trained using BN, there is no significant difference from the randomly initialized model's output (BN, rand init). BN is limited to reduce each frequency bin's deviation because BN equally normalizes all frequency dimensions.

On the other hand, when SSN is applied, the results confirm that each sub-band is independently normalized. The green line, SSN($S$=16, $A$=All), shows the effect of remarkably mitigating the scale deviation between sub-bands. When performing affine transformation for each sub-band, SSN($S$=16, $A$=Sub), our method can control a specific band's scale. It also has the effect of embedding frequency information to each sub-band.

\section{Conclusion}
\label{sec:conclusion}
In this paper, we propose a novel normalization method, SubSpectral Normalization (SSN), for the frequency domain audio input.  
SSN divides the frequency dimension into sub-bands and normalizes each of them. 
It can remove the weight deviation between sub-frequency groups while providing frequency-aware characteristics.
By changing the existing normalization layer to SSN, the user can improve the model's performance without complex model design.

\vfill\pagebreak

\bibliographystyle{IEEEbib}
\bibliography{strings,refs}

\end{document}